\documentclass[twoside,fleqn]{article}
\usepackage{landau}
\theoremstyle{definition}

\theoremstyle{remark}

\numberwithin{equation}{section}

\def\shorttitle{Title Name}
\def\shortauthor{Author 1, Author 2}

\newcommand{\subjclass}[1]{}
\newcommand{\keywords}[1]{}
\usepackage{bbm} 

\def\B{\mathbf{B}}
\def\Div{\nabla\!\cdot\!}
\def\Energy{\mathbb{E}}
\def\gasenergy{\mathcal{E}}
\def\E{\mathbf{E}}

\def\J{\mathbf{J}}
\def\Pressure{\mathbb{P}}
\def\Qf{Q^f}
\def\Qt{Q^t}
\def\qq{\mathbbm{q}}
\def\QQ{\mathbb{Q}}
\def\QQs{\mathbb{R}}
\def\QQf{\QQ^f}
\def\QQt{\QQ^t}
\def\RR{\mathbb{R}}
\def\R{\mathbf{R}}
\def\Sym{\,\mathrm{Sym}\,}
\def\curl{\nabla\!\times\!}
\def\idtens{\mathbb{I}}
\def\tr{\mathrm{tr}\,}
\def\qqq{\mathbbm{q}}
\def\q{\mathbf{q}}
\def\u{\mathbf{u}}
\def\v{\mathbf{v}}
\def\a{\mathbf{a}}
\def\viscosity{\mu}
\def\viscStress{\underline{\underline{\sigma}}}
\def\x{\mathbf{x}}
\def\s{\mathrm{s}}
\def\p{\mathrm{p}}
\def\i{\mathrm{i}}
\def\e{\mathrm{i}}
\def\tdot{\!\cdot\!}

\begin{document}

%
%
%

\def\shorttitle{Ten-moment two-fluid reconnection agrees well with PIC/Vlasov}
\def\shortauthor{E.A. Johnson, J.A. Rossmanith}

\title
{\Large \bf \boldmath Ten-moment two-fluid plasma model agrees well
with PIC/Vlasov in GEM problem\footnote{This work is supported
in part by NSF Grant DMS-0711885.}}

\author{\large E.A. Johnson
    \\ \normalsize\emph{Department of Mathematics, University of Wisconsin,}
    \\ \normalsize\emph{Madison WI 53717, USA.}
    \\ \normalsize\emph{E-mail: ejohnson@math.wisc.edu}
\\[2mm]
    \large J.A. Rossmanith
    \\ \normalsize\emph{Department of Mathematics, University of Wisconsin,}
    \\ \normalsize\emph{Madison WI 53717, USA.}
    \\ \normalsize\emph{E-mail: rossmani@math.wisc.edu}
}

\date{\vspace{-12mm}}

\maketitle

\thispagestyle{first}

\begin{abstract}
We simulate magnetic reconnection in the GEM
problem using a two-fluid model with 10 moments for the electron
fluid as well as the proton fluid. We show that use of 10
moments for electrons gives good qualitative agreement with the
the electron pressure tensor components in published kinetic
simulations.

\end{abstract}

\maketitle

\section{Overview}
This study is motivated by the following question:
\textbf{What is the simplest fluid model that can accurately
replicate kinetic (PIC and Vlasov) simulations of
fast magnetic reconnection in two-species collisionless plasma?}
Background for this question follows.

A plasma is a gas of charged particles interacting with
an electromagnetic field.  Simulations of plasma
use a variety of models of plasma, which vary greatly
in computational expense.  For a given problem one
seeks the computationally cheapest model that captures
the phenomena of interest.  We are specifically interested in
the phenomenon of fast magnetic reconnection in
``collisionless'' (i.e., low-collision) plasma.

We discuss the following sequence of plasma models
for two-species (e.g.\ electron/proton) plasmas:
   \textbf{kinetic}: Vlasov/Boltzmann or PIC (particle-in-cell),
   \textbf{ten-moment} two-fluid plasma,
   \textbf{five-moment} two-fluid plasma, and
   \textbf{MHD} (magnetohydrodynamics, a one-fluid model).
Each model in this sequence
can be regarded as a simplifying appoximation of its predecessor.
The plasma community has used particle-in-cell (PIC)
simulations as its standard first-principles plasma model.
It is an approximation of the Vlasov/Boltzmann model, which
we take as the ``truth''.
The Vlasov model is the collisionless version of the Boltzmann
model. Collisionless versions of each model are hyperbolic and
conserve entropy for smooth solutions, whereas collisional
versions are diffusive and produce entropy.

The Geospace Environmental Modeling (GEM) Magnetic
Reconnection Challenge is a benchmark magnetic reconnection
problem.  It was formulated to test the ability of plasma
models to resolve fast magnetic reconnection in collisionless
plasma.  The initial GEM studies showed that PIC simulations
exhibit fast reconnection, whereas reconnection is
much slower in MHD (with simple resistivity) \cite{article:GEM}.

\section{Plasma Models}

We implemented five-moment and ten-moment two-fluid models and
studied their ability to match published results of PIC and
Vlasov simulations.
Two-fluid models use gas-dynamics for each charged species.
These fluids are coupled to Maxwell's equations by source terms.
Hakim, Loverich, and Shumlak simulated the GEM problem with
a five-moment two-fluid model \cite{article:LoHaSh10, article:HaLoSh06} and
Hakim simulated the GEM problem with a two-fluid model
using 10 moments for ions and 5 moments for electrons \cite{article:Hakim07},
but we are unaware of any previous studies that simulate the GEM
problem with a 10-moment electron fluid.

\textbf{Boltzmann/Vlasov model.}
The Boltzmann equation asserts conservation of particle
number density $f_\s(\x,\tilde\v,t)$ in phase space:
\begin{gather*}
  \partial_t f_\s + \nabla_\x\cdot(\v f_\s)
  + \nabla_{\tilde\v}\cdot\left(\a f_\s\right)=C_\s;
\end{gather*}
here
$\tilde\v=\gamma\v\approx\v$ is (proper) velocity,
(where $\gamma=\sqrt{1+(\tilde v/c)^2}\approx 1$ is the Lorentz factor),
$\a=\frac{q_\s}{m_\s}(\E+\v\times\B)$
is the acceleration due to the electric field $\E$ and the
magnetic field $\B$,
and $C_\s$ is a collision
operator which operates on the function
$(\tilde \v,\p) \mapsto f_\p(t,\x,\tilde \v)$,
where $\p$ ranges over all species.
The Vlasov equation (collisionless Boltzmann equation)
asserts that $C_\s=0$.  The relations
$\J = \sum_\s q_\s \int_\v f_\s \v$ and
$\sigma = \sum_\s q_\s \int_\v f_\s$
couple the Boltzmann equation to Maxwell's equations
\begin{align*}
   &\partial_t \B = -\curl \E,                 & \Div\B&=0,
\\ &\partial_t \E = c^2\curl B - \J/\epsilon_0, & \Div\E&=\sigma/\epsilon_0.
\end{align*}

\textbf{Five-moment model.}
Generic physical equations for the gas-dynamic portion
of the five-moment two-fluid model consist of
conservation of mass and balance of momentum and energy for each species:
\begin{gather*}
   \!\!\!\!\!\!\!\!\!\!
   \partial_t \rho_\s + \Div(\rho_\s\u_\s) = 0, \\
   \!\!\!\!\!\!\!\!\!\!
   \partial_t (\rho_\s\u_\s) + \Div(\rho_\s\u_\s\u_\s) + \nabla p_\s
     = {q_\s\over m_\s}\rho_\s (\E + \u_\s\times\B) + \R_\s + \Div\viscStress_\s,
    \\
   \!\!\!\!\!\!\!\!\!\!
    \partial_t \gasenergy_\s + \Div(\u_\s(\gasenergy_\s+p_\s))
      + \Div\q_\s 
      = \Div(\viscStress\tdot\u) + \J_\s\tdot\E + \u_\s\tdot\R_\s
        + \Qf_\s + \Qt_\s,
\end{gather*}
where $\s$ is the species index ($\i$ for ions, $\e$ for electrons),
$\rho_\s$ is mass density,
$\u_\s$ is fluid velocity,
$\gasenergy_\s$ is gas-dynamic energy,
$q_\s=\pm \textrm{e}$ is particle charge,
$m_\s$ is particle mass, and
the species current is $\J_\s:=(q_\s/m_\s)\rho_\s \u_\s$.
For the pressure we assumed an isotropic monatomic gas:
$(3/2)p_\s = \left(\gasenergy_\s - \rho_\s u_\s^2/2\right).$
A linear isotropic entropy-respecting viscous stress closure is
$\viscStress_\s=2\mu_\s\left(\Sym(\nabla\u_\s)-\Div\u_\s\idtens/3\right)$,
where $\Sym$ denotes the symmetric part of its argument tensor,
$\idtens$ is the identity tensor, and $\mu_\s$ is the shear viscosity.
In these five-moment simulations, however,
we neglect all collisional effects. So we neglect viscosity ($\viscStress_\s=0$),
heat flux ($\q_\s=0$), resistive drag force ($\R_\s = 0$),
resistive heating ($\Qf_\s=0$), and interspecies thermal equilibration
($\Qt_\s=0$).
To couple these equations to Maxwell's equations we use the relations
\begin{align*}
   \J &= \J_i + \J_e,
   &\sigma &= \sigma_i + \sigma_e,
\end{align*}
where $\sigma_\s = (q_\s/m_\s)\rho_\s$ is the charge density
of each species.

\textbf{Ten-moment model.}
Generic physical equations for the gas-dynamic portion of the
ten-moment two-fluid model consist of
conservation of mass and balance of momentum and energy tensor
for each species:
\begin{gather*}
   \partial_t \rho_\s + \Div(\rho_\s\u_\s) = 0, \\
   \partial_t (\rho_\s\u_\s) + \Div(\rho_\s\u_\s\u_\s + \Pressure_\s)
     = {q_\s\over m_\s}\rho_\s (\E + \u_\s\times\B) + \R_\s, \\
    \partial_t \Energy_\s + 3\Div\Sym(\u_\s \Energy_\s) -2\Div(\rho_\s\u_\s\u_\s\u_\s)
      + \Div\qq_\s
      \\ \phantom{sp}
      = {q_\s\over m_\s}2\Sym(\rho_\s\u_\s\E+\Energy_\s\times\B) + \QQs_\s
        + 2\Sym(\u_\s\R_\s) + \QQf_\s + \QQt_\s,
\end{gather*}
where $\Energy_\s:=\int_\v f_\s \v\v$ is the energy tensor and
$\Pressure_\s:=\Energy_\s-\rho_\s\u_\s\u_\s$ is the pressure tensor.
A linear isotropic entropy-respecting isotropization closure is
$\RR_\s = \frac{1}{\tau_\s}\left({1 \over 3}(\tr\Pressure_\s) \idtens - \Pressure_\s\right),$
where $\tr$ denotes the trace of its argument tensor
and for the isotropization period we used
$\tau_\s = \tau_0 \sqrt{\frac{\det\Pressure_\s}{\rho_\s^5}}m_\s^3,$
which attempts to generalize the Braginskii closure; for the GEM problem
this means that $\tau_i/\tau_e \approx (m_i/m_e)^{5/4}$.
We neglect all other collisional terms: the heat flux tensors $\qqq_\s$,
the resistive drag forces $\R_\s$, the frictional heating tensors $\QQf_\s$,
and the temperature equilibration tensors $\QQt_\s$.

For small viscosity (or fast isotropization)
the viscosity is related to the isotropization period by
$\viscosity_\s \approx p_\s\tau_\s$.  We set $\tau_0=50$.
The ten-moment model offers the advantage of
\emph{hyperbolic viscosity} --- that is, viscosity can
be implemented without numerically expensive diffusive terms.

\section{GEM Problem}

The GEM problem is posed on a rectangular domain
with periodic boundary conditions in the horizontal direction
and with conducting wall boundary conditions for
the upper and lower boundaries.
The initial conditions are a Harris sheet equilibrium
perturbed by ``pinching'' to form an X-point.

\def\GEM{\mathrm{GEM}}

\textbf{Nondimensionalization.}
The GEM problem nondimensionalizes
time by the ion gyrofrequency
$\Omega_i=\frac{e B_0}{m_\i}$
and the space scale by the ion inertial length
$\delta_i$, which is the distance traveled in an ion
gyroperiod $1/\Omega_i$ at the ion 
Alfv\'en speed $v_{A,\i}:=\frac{B_0}{\mu_0 m_\i n_0}$
(where $\mu_0:=(c^2\epsilon_0)^{-1}$ is magnetic permeability).
Under this nondimensionalization the model equations above
remain unchanged with the exception that $\epsilon_0$
is replaced by $\epsilon:=1/c^2$
(where $c$ is now the speed of light divided by the
ion Alfv\'en speed).


\textbf{Model Parameters.}
The GEM problem specifies that the ion/electron mass
ratio is $m_i/m_e=25$ and the initial temperature ratio is 
$T_i/T_e = \sqrt{m_i/m_e}=5$.
The speed of light is not specified; we used the
commonly used value of $20$, chosen to be sufficiently high
to exceed other wave speeds but computationally feasible.

\textbf{Computational domain.}
The computational domain is the rectangular domain
$[-L_x/2,L_x/2]\times[-L_y/2,L_y/2]$,
where $L_x=8\pi$ and $L_y=4\pi$.
The problem is symmetric under reflection across
either the horizontal or vertical axis.

\textbf{Boundary conditions.}
The domain is periodic in the $x$-axis.
The boundaries perpendicular to the $y$-axis
are thermally insulating conducting wall boundaries.  
A conducting wall boundary is a solid wall boundary
(with slip boundary conditions in the case of ideal plasma)
for the fluid variables, and the electric field at
the boundary has no component parallel to the boundary.
We also assume that magnetic field runs parallel to
and so does not penetrate the boundary (this follows
from Ohm's law of ideal MHD, but we assume it holds
generally).  So at the conducting wall boundaries
\begin{align*}
  & \partial_y \rho_\s = 0,
  &  
    \partial_y B_x = 
    B_y = 
    \partial_y B_z = 0,
  \\
  & \partial_y u_{\s x} = 
    u_{\s y} = 
    \partial_y u_{\s z} = 0,
  &  
    E_x = 
    \partial_y E_y = 
    E_z = 0.
\end{align*}

\textbf{Initial conditions.}
The initial conditions are a perturbed Harris sheet equilibrium.
The unperturbed equilibrium is given by
\def\e{\mathbf{e}}
\def\E{\mathbf{E}}
\def\J{\mathbf{J}}
\def\sech{\,\mathrm{sech}}
\begin{align*}
    \B(y) & =B_0\tanh(y/\lambda)\e_x,
  & p(y) &= \frac{B_0^2}{2 n_0} n(y),
 \\ n_i(y) &= n_e(y)
            = n_0(1/5+\sech^2(y/\lambda)),
  & p_e(y) &= \frac{T_e}{T_i+T_e}p(y),
 \\ \E & =0,
  & p_i(y) &= \frac{T_i}{T_i+T_e}p(y).
\end{align*}
On top of this the magnetic field is perturbed by
\begin{align*}
   \delta\B&=-\e_z\times\nabla(\psi), \hbox{ where}
\\ \psi(x,y)&=\psi_0 \cos(2\pi x/L_x) \cos(\pi y/L_y).
\end{align*}
In the GEM problem the initial condition constants are
\begin{align*}
    \lambda&=0.5,
  & B_0&=1,
  & n_0&=1,
  & \psi_0&=B_0/10.
\end{align*}


\section{Method}
To simulate the ten-moment and five-moment systems
we implemented a Runge-Kutta
Discontinuous Galerkin solver with third-order accuracy
in space and time on a Cartesian mesh.

To suppress oscillations, after each time stage we limited
the solution in the characteristic variables of the cell
average using a modification of Krivodonova's method
(beginning with the coefficients of the
highest-order Legendre basis polynomials and descending
to lower order if limiting occurs) \cite{article:Krivodonova07}.

\def\diminished{}
\def\emphasized{}
To clean the magnetic field we
used a correction potential ${\diminished \psi}$ in Maxwell's equations,
as suggested in \cite{article:DeKeKrMuScWe01}:
\begin{gather*}
 \!\!\!\!\!\!\!
 \partial_t \B + \curl\E{\diminished +\chi\nabla\psi} = 0, 
 \ \ 
 \partial_{t} \E - c^2\curl \B = -\J/{\epsilon_0}, 
 \ \ 
 {\diminished \partial_{ t} {\diminished\psi} + }
 {\diminished \chi c^2{\emphasized  \Div\B}} = {\emphasized 0}.
\end{gather*}
These equations imply a wave equation that propagates
the divergence constraint error $\Div\B$ at the speed $c\chi$.
We used $\chi=1.05$.

Since the GEM problem is symmetric,
we imposed symmetry and solved the equations on the quarter domain
$[0,L_x/2]\times[0,L_y/2]$.

\section{Results}

We simulated the GEM problem with ten-moment and five-moment models
and compared the results with the Vlasov simulations of \cite{article:ScGr06}
and the PIC simulations of \cite{article:Pritchett01}.\footnote{
We have to negate some quantities because we call the vertical
axis $y$ and the out-of-plane axis $z$, opposite to the
convention of \cite{article:Pritchett01, article:ScGr06}.}
Their plots were made at a point in time when the
flux through the positive $x$-axis approximately reaches
1 nondimensionalized unit.\footnote{
The initial flux through the positive $x$-axis is $2\psi_0=0.2$
and the initial total flux through the positive $y-$axis
is $\ln(\cosh(2\pi))-(0.1(=\psi_0))\approx 5.4900$.
So the percent reconnection at this time is
$0.8/5.49 \approx 14.6\%$.
} This is shortly before the time when
the reconnection rate peaks.


As measured against kinetic simulations
the ten-moment model reconnects at about the correct rate
and the five-moment model reconnects a bit too quickly,
perhaps because in the five-moment model compression in the outflow direction
automatically causes increased pressure in the perpendicular
directions, opening up the outflow region and artificially
increasing the rate of reconnection.

In contrast to the five-moment model the ten-moment model 
is capable of representing an anisotropic pressure tensor.
Our plots of the ten-moment electron pressure tensor components
look like somewhat smudged versions of the corresponding
plots for the Vlasov model.


 \begin{figure}[t!]
   \includegraphics[height=35mm,width=60mm]{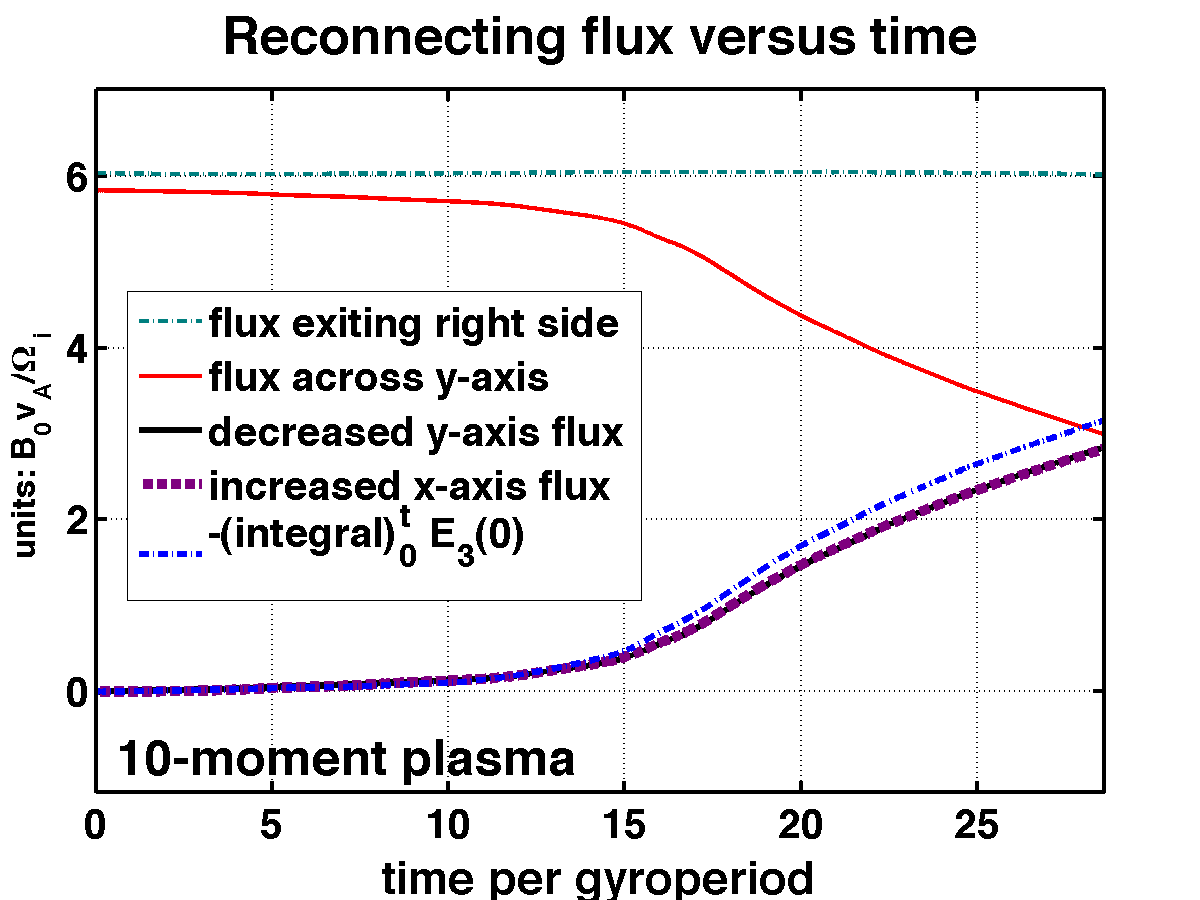}
   \includegraphics[height=35mm,width=60mm]{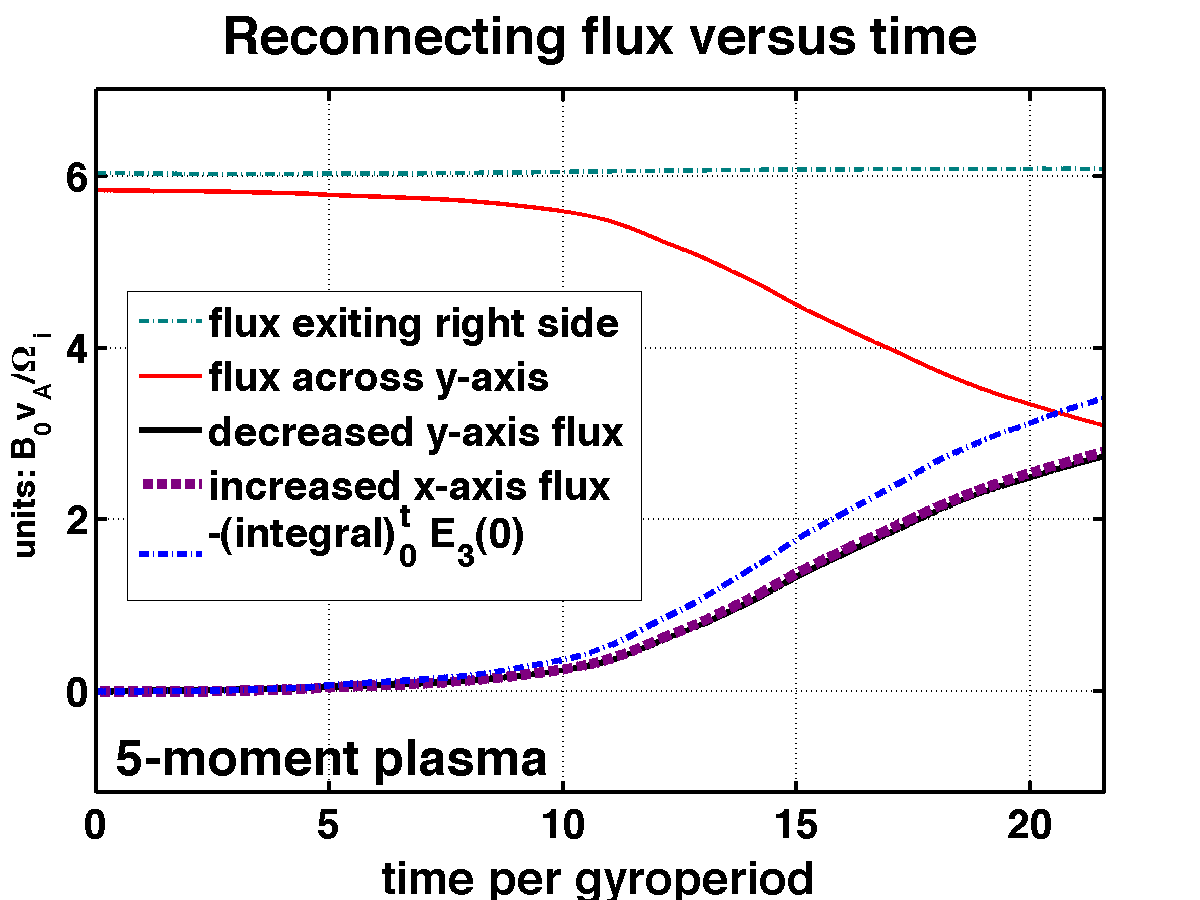}
 \caption{
   Ramp-up of reconnected flux for 10-moment and 5-moment plasma.
   Fluxes are through the boundaries of the first quadrant
   of the domain.  In each model
   the bottom three lines should coincide.
   Times at which the magnetic flux through the positive $x$-axis
   reached one nondimensionalized unit per model were
   $17.2$ (our 10-moment) 
   $13.0$ (our 5-moment),
   $17.7$ (Vlasov \cite{article:ScGr06}),
   $15.7$ (PIC \cite{article:Pritchett01}),
   $17.6$ (10/5-moment \cite{article:Hakim07}),
   $15.6$ (5-moment \cite{article:HaLoSh06}) and
   $15.3$ (5-moment \cite{article:LoHaSh10}).
  }

 \label{f:recon}
 \end{figure}


\begin{figure}[t!]

  \includegraphics[height=40mm,width=60mm]{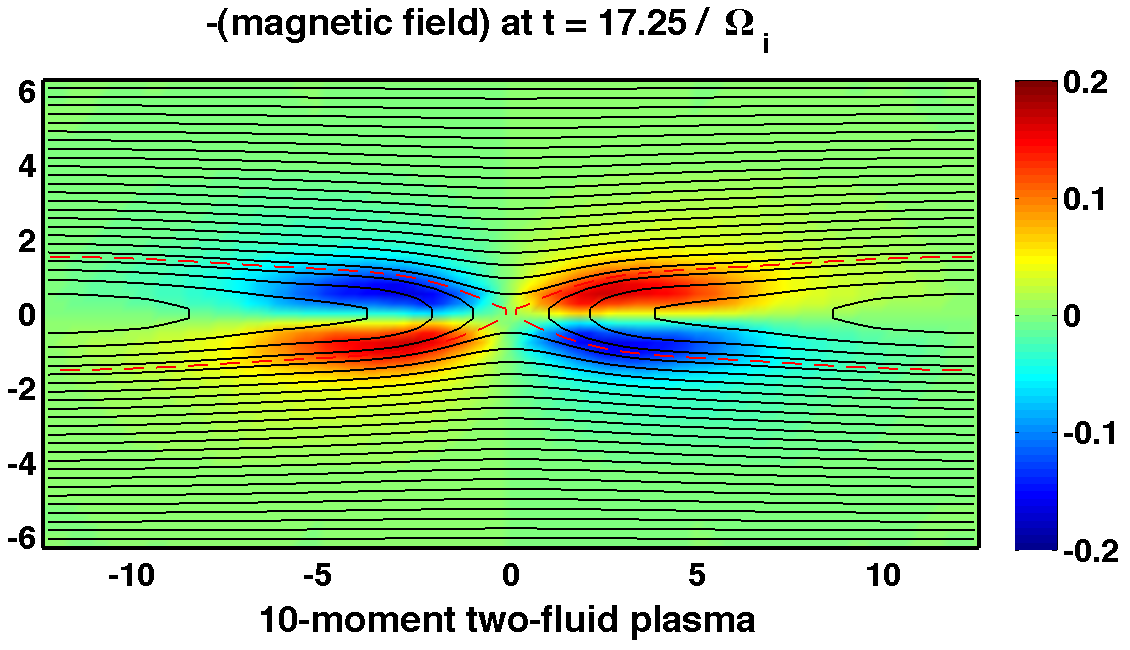}
  \includegraphics[height=40mm,width=60mm]{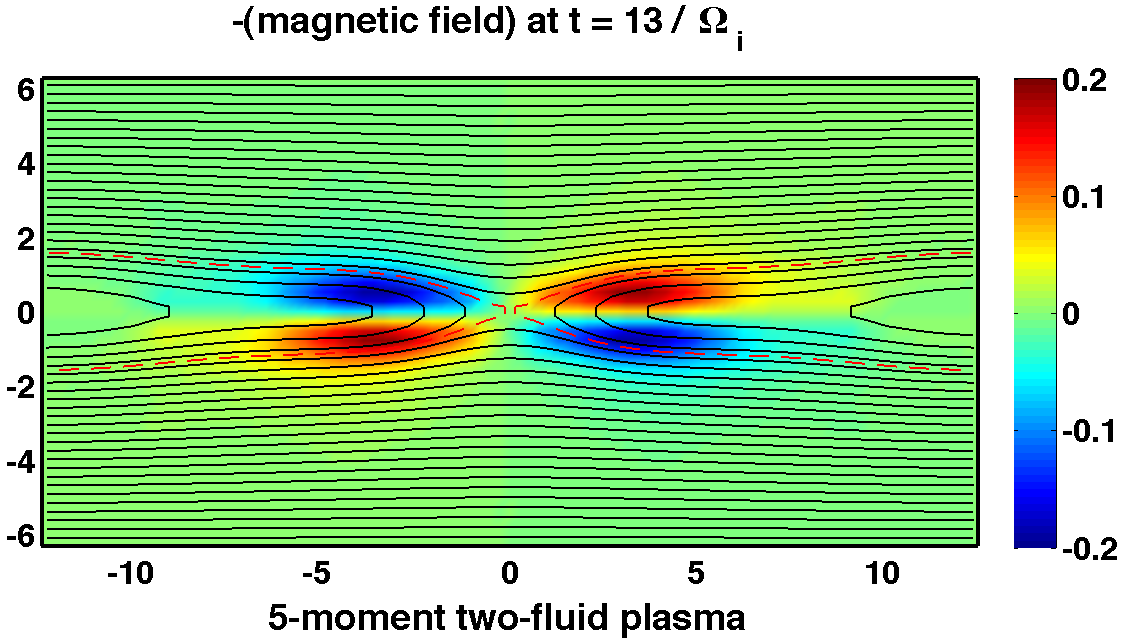}
  \caption{
    Magnetic field when one unit of flux passes through
    the positive $x$-axis.  These results compare well with 
    Plate 1 of \cite{article:Pritchett01} and the magnetic
    field shown in FIG. 2 of \cite{article:ScGr06}.
  }
  \label{f:B2}
\end{figure}


\begin{figure}[t!]
    \includegraphics[height=25mm,width=120mm]{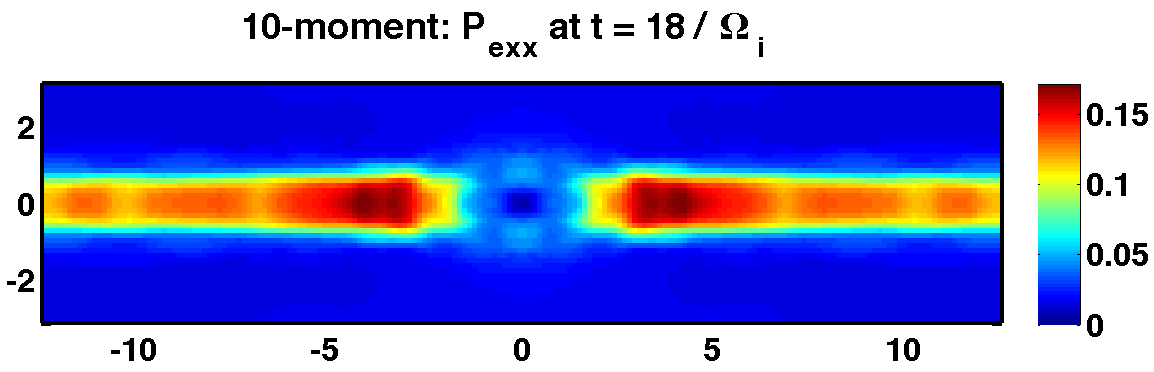}
 \\ \includegraphics[height=25mm,width=120mm]{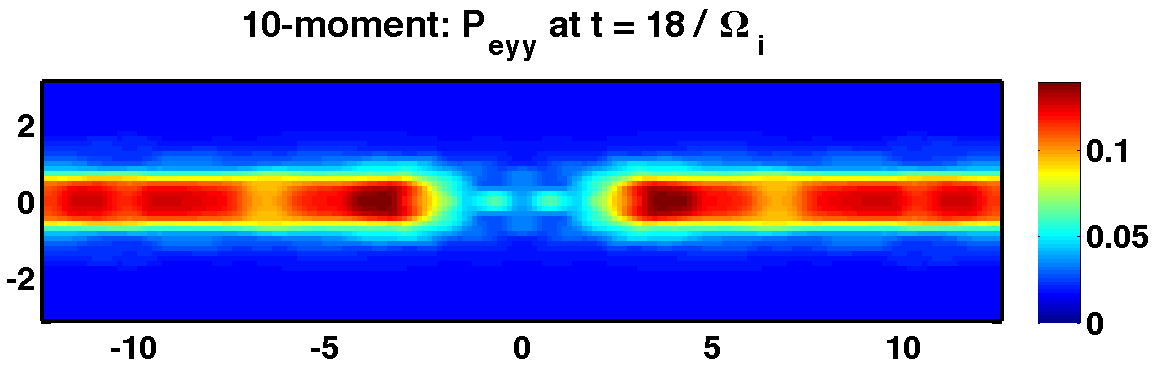}
 \\ \includegraphics[height=25mm,width=120mm]{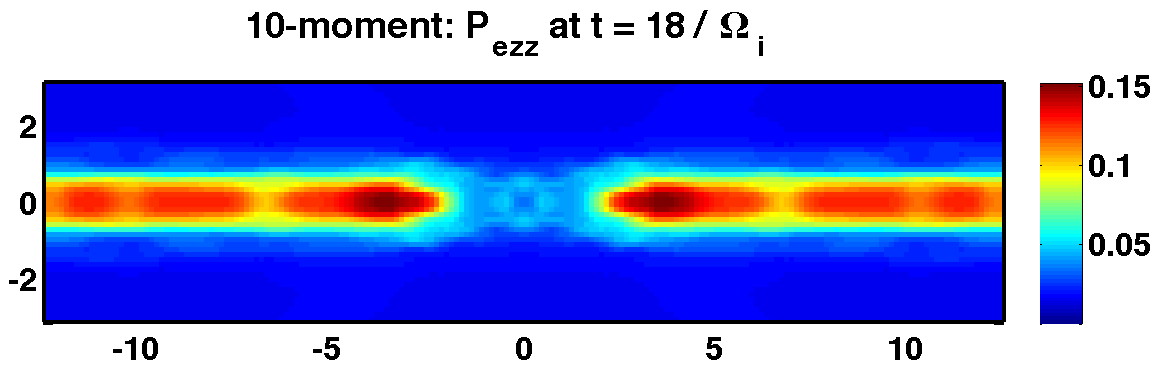}
 \caption{
  Diagonal components of the electron pressure tensor
    for 10-moment simulation at $\Omega_i t = 18$.
    These look like smudged versions of the plots
    in FIG. 5 of \cite{article:ScGr06}.
    }
\end{figure}

\begin{figure}[t!]
    \includegraphics[height=25mm,width=120mm]{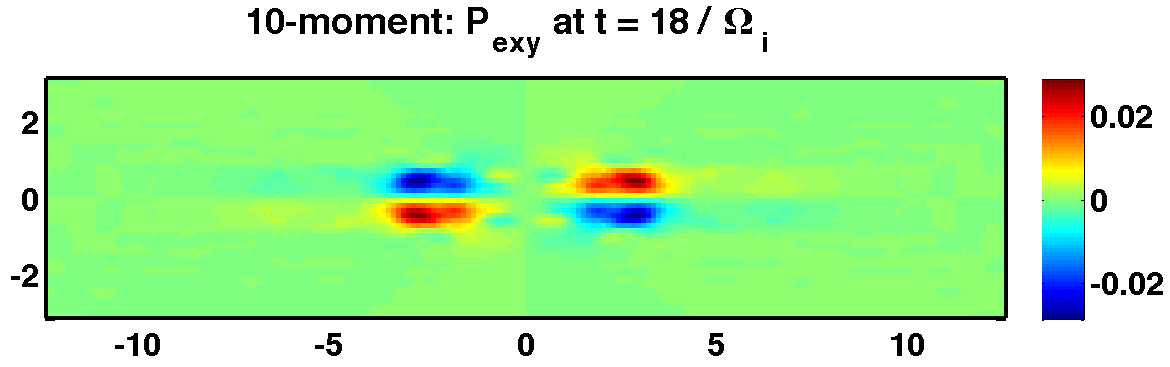}
 \\ \includegraphics[height=25mm,width=120mm]{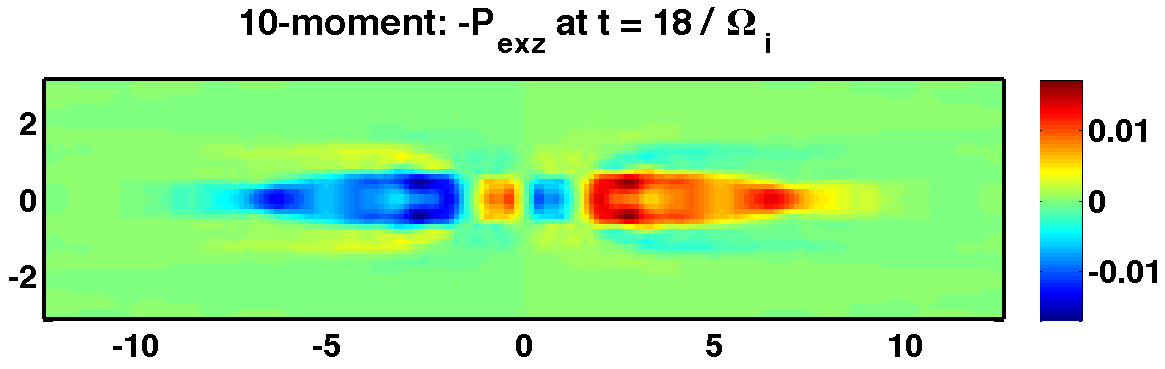}
 \\ \includegraphics[height=25mm,width=120mm]{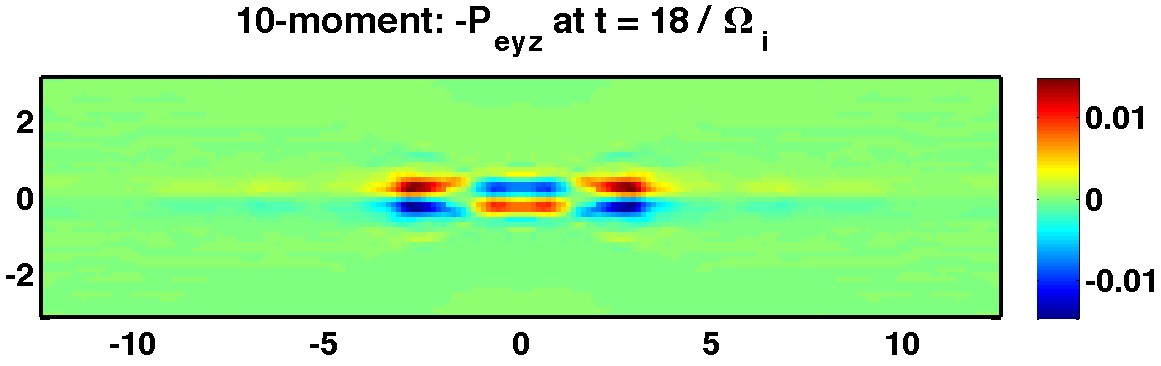}
 \caption{
  Off-diagonal components of the electron pressure tensor
    for 10-moment simulation at $\Omega_i t = 18$.
    These look like smudged versions of the plots in
    FIG. 6 of \cite{article:ScGr06}.
    }
\end{figure}


\end{document}